\documentclass{article}
\usepackage{arxiv}

\usepackage[utf8]{inputenc} 
\usepackage[T1]{fontenc}    
\usepackage{hyperref}       
\usepackage{url}            
\usepackage{booktabs}       
\usepackage{amsfonts}       
\usepackage{nicefrac}       
\usepackage{microtype}      
\usepackage{lipsum}

\usepackage[round]{natbib}

\usepackage{amsfonts}       
\usepackage{nicefrac}       

\usepackage{algorithm}
\usepackage[noend]{algpseudocode}

\usepackage{tabularx}
\usepackage{multirow}

\usepackage{amsfonts}
\usepackage{amsmath}
\usepackage{amsthm}
\usepackage{mathtools}
\usepackage{bm}
\usepackage{bbm}

\usepackage{caption}
\usepackage{algorithm}
\usepackage[noend]{algpseudocode}
\usepackage{amssymb}
\usepackage{mathtools}
\usepackage{amsmath}
\usepackage{tikz}
\usetikzlibrary{calc}
\usepackage{todonotes}
\usepackage{dsfont}
\usepackage{enumitem}
\usepackage{tikz}
\usepackage{amsthm}
\usepackage{thm-restate}
\usepackage{booktabs}
\usepackage{multirow}
\usepackage{newfloat}
\usepackage{wrapfig}
\usepackage{comment}
\usepackage{multirow}
\usepackage{subfig}
\usepackage{tabu}

\newcommand{\alg}{\textsf{PPTO}}

\usepackage[switch]{lineno}
\makeatletter
\AtBeginDocument{%
	\@ifpackageloaded{amsmath}{%
		\newcommand*\patchAmsMathEnvironmentForLineno[1]{%
			\expandafter\let\csname old#1\expandafter\endcsname\csname #1\endcsname
			\expandafter\let\csname oldend#1\expandafter\endcsname\csname end#1\endcsname
			\renewenvironment{#1}%
			{\linenomath\csname old#1\endcsname}%
			{\csname oldend#1\endcsname\endlinenomath}%
		}%
		\newcommand*\patchBothAmsMathEnvironmentsForLineno[1]{%
			\patchAmsMathEnvironmentForLineno{#1}%
			\patchAmsMathEnvironmentForLineno{#1*}%
		}%
		\patchBothAmsMathEnvironmentsForLineno{equation}%
		\patchBothAmsMathEnvironmentsForLineno{align}%
		\patchBothAmsMathEnvironmentsForLineno{flalign}%
		\patchBothAmsMathEnvironmentsForLineno{alignat}%
		\patchBothAmsMathEnvironmentsForLineno{gather}%
		\patchBothAmsMathEnvironmentsForLineno{multline}%
	}{}
}
\makeatother

\title{A privacy-preserving tests optimization algorithm for epidemics containment}

\author{
	Alessandro~Nuara\\
	Politecnico di Milano\\
	\texttt{alessandro.nuara@polimi.it}
	\And
	Francesco~Trov\`o \\
	Politecnico di Mil1ano\\
	\texttt{francesco1.trovo@polimi.it}
	\And
	Nicola Gatti\\
	Politecnico di Milano\\
	\texttt{nicola.gatti@polimi.it}
}

\begin{document}

\maketitle



\begin{abstract}
The SARS-CoV-2 outbreak will affect the everyday life of billions of people worldwide for a long period.
There is a common agreement that the safe reactivation of the production processes after the health emergency, necessary to prevent economic and social crises, cannot be delivered without an effective contact tracing activity. 
In particular, two classes of people, namely \emph{asymptomatic} and \emph{presymptomatic} individuals, play a crucial role in the spread of the contagion and should be traced promptly. 
The former is an important fraction of the infected people and their identification is difficult, while the latter can transmit the virus with high probability. 
Contact-tracing applications can generate millions of punctual data that can be used to reconstruct infection chains and predict future infections, but their processing is intractable without automatic techniques such as Artificial Intelligence (AI). 
Furthermore, many countries adopted severe privacy regulations forcing the information on each individual to be anonymized and resides on her personal device only, thus potentially limiting the capabilities of AI algorithms. 
In this paper, we design a novel algorithm, namely \alg{}, that reconstructs \emph{forward} and \emph{backward infection chains} in a distributed fashion, exploiting the communication of anonymized information among the personal devices and, thus, preserving the privacy. 
Our algorithm also solves, in a distributed fashion, an online optimization problem to identify the individuals to be tested and potentially isolated in the attempt to minimize the spread of the contagion.
We provide an experimental evaluation of our algorithm in synthetic settings.
\end{abstract}

\section{Introduction}

The health emergency is the first challenge each country has faced in the fight to SARS-CoV-2.
The next challenge, even more critical, is the safe reactivation of the production processes to prevent economic and social crises.
Even if the vaccine should be successful, its distribution to the whole population will require more than one year, thus forcing a long coexistence with the virus.
During this time, it will be crucial to understand how the infection is spreading across the population and take the best countermeasures to contain it without further lockdowns.
The epidemic models currently used provide a macro-scale description of the spread, which allows forecasting the hospitals' load during the lockdown, but are partially ineffective when preventing the lockdown.
An effective tool to analyze the current virus spread situations and identify the presence of infected individuals is the use of swabs, which are used extensively. 
However, the limited availability and the time required to test a large number of subjects make them only a partial answer to the reconstruction of the disease transmission among the population.

The scientific community agrees that contact-tracing applications could provide micro-scale data useful to contain the infection by driving the choice of the daily swab selection process.
These applications can generate millions of punctual data that can be combined with the clinical data available to the Local Health Units (LHUs) to reconstruct infection chains and networks and, thus, identify potentially infected people.\footnote{
For instance, the contact-tracing app adopted in Italy is \url{https://www.immuni.italia.it/}.}
Currently, only direct contacts with the positive subjects are tested for SARS-CoV-2, while a more accurate analysis can suggest testing those subjects which have a high probability of being infected even if they had not direct contact with the positive one.
However, dealing with millions of data is unfeasible for humans and Artificial Intelligence (AI) techniques are commonly considered the most promising technology for such tasks.
Remarkably, \citet{tsai2012security} have already adopted AI techniques for the contagion containment problem in a setting different from ours (\emph{i.e.}, fake news containment in social networks).
In the SARS-CoV-2 setting, AI can be adopted to develop predictive models from data and reasoning on them to suggest where and when to use swabs to test for infected people. 
However, the potential of these techniques is not fully understood yet. 
More importantly, their applicability is limited by the severe European privacy regulation that requires the information on the individuals to only reside on their personal devices in an anonymized fashion.
A natural choice to address this problem is the adoption of distributed algorithms capable of managing contacts information in a decentralized fashion~\cite{DBLP:conf/atal/AliKT05,DBLP:conf/ijcai/WuZJ13}. 
Distributed algorithms are extensively studied in the multi-agent systems literature, where every agent locally controls a subset of variables and exchanges messages with the other agents to achieve a solution (\citet{shoham2008multiagent}). 
%
%
%

\paragraph{Original Contributions} In this work, we present a distributed algorithm, namely \alg, capable, in a decentralized fashion as to avoid the violation of the privacy regulation, of identifying two classes of people whose roles are prominent in the infection from contact-tracing data: \emph{asymptomatic} subjects, \emph{i.e.}, infected subjects not showing any symptom, and \emph{presymptomatic} subjects, \emph{i.e.}, individuals in the initial phase of the disease.
The former ones are an important fraction of the infected people, and their identification is difficult, while the latter can transmit the disease with high probability.
%
%
We develop a micro-scale epidemic model based on the recent virus transmission studies by~\citet{ferretti2020quantifying}.
The \alg{} algorithm exploits this model to reconstruct infection chains present in the population using tracing app data and ranks the individuals according to the likelihood of being infected, so that the limited amount of daily swab tests available to the healthcare authorities can be used effectively.
Finally, we experimentally compare on a synthetic setting the \alg{} algorithm with currently adopted policies to select swabs.

\section{Related Works}

%
Many efforts have been spent in the modeling of SARS-CoV-2 disease by means of predictive approaches depending on the specific areas of the virus diffusion, \emph{e.g.,}~\citet{girardi2020robust} and~\citet{gaeta2020data} for Italy, \citet{toubiana2020estimated} for France, and~\citet{chen2020two} for China.
%
%
Instead, \citet{staszkiewicz2020dynamics} provide a broader study to identify the main factors affecting the SARS-CoV-2 virus spread, showing that the spread and mortality are determined by the number of connections among the continents (\emph{e.g.}, the airline traffic) and by some local characteristics of the country (\emph{e.g.}, medical personnel and average air pollution).
Even if the previous studies provide mathematical epidemic models describing the macro evolution of a generic disease, which allow to forecast the hospitals' load during the lockdown, they are partially ineffective to prevent a new lockdown.
Most importantly, they do not provide any support to design the best strategies to choose the individuals to test every day.
%
Models describing the spread of contagion are customarily adopted in the AI literature, \emph{e.g.}, for the social influence maximization problem where the goal, different from ours, is to maximize the spread of information over a social network, modeled as a directed graph, by selecting a set of optimal nodes representing influential individuals~\cite{wen2017online,peng2019adaptive,kamarthi2020influence}.
These approaches can be also adapted to model the spread of a disease, where the contacts are modeled as edges in a graph, in which each individual (or group of individuals) is  represented by a node.
%
%
\citet{prasse2020network} study the evolution of the disease over the network, using a large-scale approximation of the contacts, in which every city is represented as a node of a graph.
These findings allows to take decisions on a macro scale, \emph{e.g.}, deciding the number of tests to assign to the different areas, but they do not provide suggestions on the single individuals to test.
Instead, the goal of the work by~\citet{shaghaghian2017online} is to infer the path that an infection has traversed in order to reach an arbitrary node in the Susceptible Infected Recovered (SIR) epidemiological model. 
Unfortunately, this approach does not provide a clear decision on the tests to perform to minimize the future SARS-CoV-2 virus spread.
Using message passing algorithms and statistical inference by maximum likelihood approach (see the work by~\citet{lokhov2014inferring,antulov2014statistical}) study the disease graph to infer the infection source. 
%
%
However, the solutions cited above are not viable in practice since they require that the information on the contacts between individuals is elaborated in a centralized manner.
This approach raises major privacy issues, since attacks to the centralized unit would compromise the information on the contacts occurred to the entire population.
To the best of out knowledge, no solution taking into account privacy issues has been proposed so far for the identification of infection sources and paths.

An technical solution to contact tracing is provided by tracing apps, which can be used to identify the individuals with the most likelihood of being infected and their selection for the SARS-CoV-2 swab test (see~\citet{ahmed2020survey} for an extensive review).
%
%
Except for some first-generation apps using directly users personal data~\cite{tang2020privacy}, they are required to implement the PACT protocol~\cite{chan2020pact}, which regulates the privacy of contact-tracing apps and has been recognized by a large number of world governments.
Using the guidelines provided by this protocol, the contact tracing apps are only allowed to exchange some identification codes between devices owned by individuals who had a prolonged contact.
From the technological point of view, Apple and Google are developing the guidelines for Bluetooth-based contact tracing apps.\footnote{\url{www.apple.com/covid19/contacttracing}.}
This allows people using these apps to preserve their personal information about the contacts they had in the past days and, at the same time, allows the authorities to determine which are the people who have been the directly exposed to infected individuals.
Nonetheless, so far no intelligent mechanisms has been included to such apps to reconstruct the infection path over time.
The method we design in this paper is based on the information gathered by such apps.
Furthermore, it is in line with the PACT protocol, and it uses an implementation in line with the Apple-Google technological guidelines.

Our work is also related to the field of decentralized optimization widely studied in multi-agent systems, in which the algorithms work in a distributed fashion.
In particular, each agent can control a subset of variables, and the agents exchange information to reach an optimal solution. 
Well-known examples are ADOPT by~\citet{DBLP:conf/atal/AliKT05} and~\citet{matsui2005efficient}, and DPOP by~\citet{inproceedingsDPOP} for optimization problems with constraints.
Two main issues distinguish our work from those in the multi-agent field: we do not need all the machinery introduced by the above works as our problem does not present constraints, and the need for privacy would limit excessively the information exchanged by the agents.

\section{Epidemic Model}

In this work, we use a new epidemic model that allows us to simulate a virus spread in a fine-grained fashion.

\paragraph{Populations}
In this model, namely \textit{SAPSR}, at each time $t \in \{1, \ldots, T \}$, the individuals of a population are partitioned into $5$ classes:
\begin{itemize}
	\item $S_t$: \textit{Susceptible} individuals that did not contract the virus before $t$;
	\item $A_t$: \textit{Asymptomatic} individuals that contracted the virus before $t$ and will not manifest any symptom during the whole development of the disease;
	\item $P_t$: \textit{Presymptomatic} individuals that contracted the virus before $t$ without manifesting any symptom, but they will manifest some symptoms in the future;
	\item $Y_t$: \textit{Symptomatic} individuals that contracted the virus before $t$ and present some symptom;
	\item $R_t$: \textit{Recovered} individuals that were either asymptomatics or symptomatics before $t$, but that at $t$ are not infected anymore. This category, at $t$, cannot infect other people and are immune to the disease.\footnote{Though recent studies showed that a few individuals can be reinfected, in this work, where we consider relatively small time horizons, we do not include these scenarios.}
\end{itemize}
The separation between asymptomatic and presymptomatic individuals in epidemic models is proposed by~\citet{ferretti2020quantifying} to capture properly the SARS-CoV-2 outbreak dynamics, as the virus transmission is mainly conveyed by these two classes of individuals.
Indeed, symptomatic people are less likely to spread the contagion since, usually, they are promptly isolated as symptoms occur, while asymptomatics and presymptomatics have frequent contacts with susceptible individuals, which, consequently, are more likely to be infected by individuals from these classes.
However, once a contact occurs, symptomatics have the largest probability to infect a new individual, followed by presymptomatics and asymptomatics, who have the lowest probability of passing the disease.
Thus, in our model, the probability of spreading the disease is differentiated among the three classes.

\begin{figure}[t!]
	\centering
	\includegraphics[width = 0.65\textwidth]{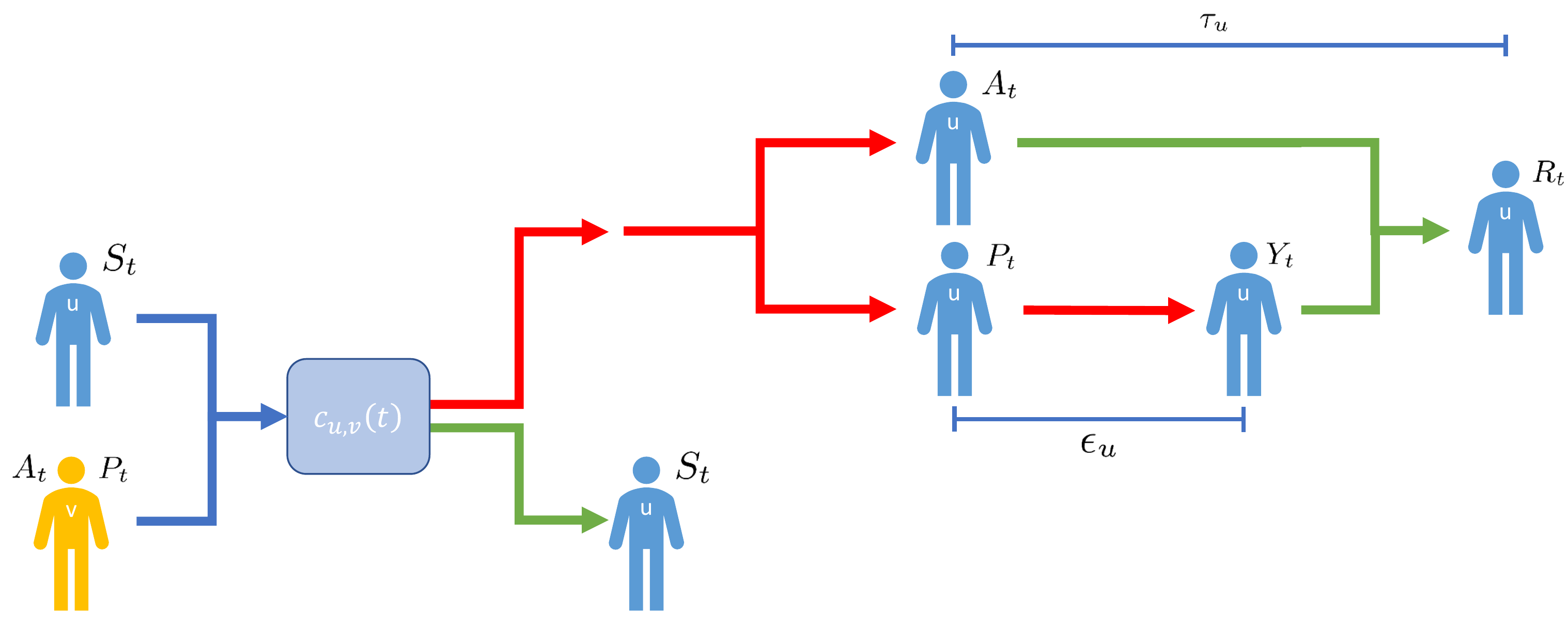}
	\caption{Contagion scheme: when a susceptible $S_t$ individual $u$ (in blue) has a contact $c_{u,v}(t)$ with an asymptomatic or presymptomatic one $v$ (in yellow), she/he may get infected. If so, she/he may become an asymptomatic $A_t$ or a presymptomatic $P_t$ patient. In the former case, after a period of $\tau_u$ she/he recovers $R_t$. Conversely, in the latter case, the presymptomatic status becomes symptomatic $Y_t$ after a period of $\epsilon_u$ days, and after a period she/he recovers $R_t$.}
	\label{img:contagion}
\end{figure}

\paragraph{Contact Model}
Our model describes contacts and contagions among individuals during a period of $T$ days.\footnote{
The quantities used in what follows are listed in Table~\ref{tab:notation}, available in the Supplementary Material.}
%
%
Let $U_t$ be the population individuals involved in the contagion.
We define with $w_{u,v}(t)$ the probability that two nodes $u \in U_t$ and $v \in U_t$ have a contact at time $t$.
We set a different value for each pair of individuals to model both individuals pairs that have recurrent and frequent contacts, \emph{e.g.}, family members, colleagues, and friends, while others have only sporadic encounters, \emph{e.g.}, people using public transportation.
Notice that in the current study we focus on the transmission of the virus by direct contact between couples of people, \emph{i.e.}, occurred by exchanging respiratory droplets, since it is known that the virus can survive only few hours outside the human body~\cite{khadka2020different}. 
However, the model can be easily extended to include also other forms of environmental contagion.

A contact between two individuals $u \in U_t$ and $v \in U_t$ is denoted with $c_{u,v}(t)$, and it is defined as a tuple $(l_{u,v}(t), r_{u,v}(t), t)$, where $l_{u,v}(t)$ denotes the distance between $u$ and $v$, $r_{u,v}(t)$ denotes the duration of the contact, and $t$ is the time the contact between $u$ and $v$ occurred.\footnote{
Even though, other parameters can be relevant to determine the contagion probability, here, we focus only on these parameters as they constitute the information provided by the most common contact-tracing apps~\cite{khadka2020different}.}
Let us denote with $C(t) = \{ c_{u,v}(t) \text{ s.t. } u, v \in U_t \}$ the set of all the contacts occurred between pairs of individuals $(u, v)$ during day $t$.
Note that $c_{u,v}(t) \in C(t)$ also implies $c_{v,u}(t) \in C(t)$, since contacts are symmetric.

\paragraph{Contagion Model}
A graphical representation of the contagion model is presented in Figure~\ref{img:contagion}.
Let us denote with $\delta(u, v, l_{u,v}(t), r_{u,v}(t)) \in [0, \ 1]$ the contagion probability between two individuals $u \in U_t$ and $v \in U_t$ who have a contact for a duration of $r_{u,v})(t)$ at a distance $l_{u,v}(t)$.\footnote{
This probability can be inferred from epidemiological models or it can be estimated resorting to data about the past individuals contacts.}
This probability is non-null only if the node $u \in S_t$, \emph{i.e.,} she/he is susceptible, and the node $v \in A_t \cup P_t \cup Y_t$, \emph{i.e.,} she/he have previously contracted the SARS-CoV-2, and is different depending on the set $v$ belongs to (see~\citet{ferretti2020quantifying} for details).
If a susceptible node $u \in S_t$ becomes infected, she/he is removed from the population $S_t$ and included into $A_t$ with probability $\alpha_s(u)$, or she/he is included in $P_t$ otherwise.
If an individual becomes asymptomatic at a specific time $t_0$, she/he will remain in the asymptomatic population $A_t$ for a time span of $t \in (t_0, t_0 + \tau_u)$, where $\tau_u$ is the time required by the asymptomatic individual $u$ to recover from the virus, \emph{i.e.,} to result not infected anymore.
Subsequently, at time $t_0 + \tau_u$, she/he will be removed from the set of the asymptomatic individuals $A_{t_0 + \tau_u}$ and included in the recovered population $R_{t_0 + \tau_u}$.
Conversely, if an individual becomes presymptomatic at time $t_0$, she/he will remain in population $P_t$ for $t \in (t_0, t_0 + \epsilon_u)$, and, at time $t + \epsilon_u$, she/he will be included in the symptomatic population $Y_{t + \epsilon_u}$, where $\epsilon_u$ denotes incubation time for individual $u$.

\paragraph{Containment Model}

While the above mechanisms describe the development of the contagion in a completely uncontrolled population, the introduction of containment measures deeply affect how the virus spreads among the population.
As containment measure, in this work, we test each day a subgroup $V_t$ of cardinality at most $K$ of the population $U_t$, and we isolate those individuals with a positive outcome.
In this way, we can exclude them from the generation of new dangerous contacts from $C(t+1)$.
%
%
%
A generic \emph{testing policy} having a maximum number $K$ swab tests of is a function that, given the contacts in the past days $C(1:t) := \cup_{h=1}^t C(h)$ and the current population $U_t$, returns a subset $V_t$ to be tested with $|V_t| = K$.
Once the results on the individuals in $V_t$ are ready, the positive ones in the set $X_t$, are isolated from the population that can freely circulate from the next day $U_{t+1}$.
Notice that if an individual has been tested at some point, this does not exclude from her/his test in the future due to the false positives occurring during the test and the possibility that the contagion occurred during the successive days.

%
%

\begin{algorithm}[t!]
\centering
\small
\caption{Epidemic development} \label{alg:env}
\begin{algorithmic}[1]
	\State Input: initial population sets $S_0$, $A_0$, $P_0$, $Y_0$, $R_0$, time horizon $T$, tests policy $\texttt{testPolicy}$, clinical test $\texttt{test}$
	\State Output: final population sets $S_T$, $A_T$, $P_T$, $Y_T$, $R_T$
	\For{$u \in U_0$}
		\State $D_u \gets \emptyset$
	\EndFor
	\For {$t \in \{1, \ldots, T \}$}
	\For{$u, v \in U_t$}
		\For{$c_{u,v}(t) \in C(t)$}
			\State $(D_u, D_v) \gets $\texttt{updateDevData}$(D_u, D_v, c_{u, v}(t))$
		\EndFor
	\EndFor
	\State $(S, A, P, Y, R_t) \gets $\texttt{contagionStep}$(C(t), S_{t-1}, A_{t-1},$ $P_{t-1}, Y_{t-1}, R_{t-1})$ \label{line:cont}
	\State $V_t \gets \texttt{testingPolicy}(C(1:t), U_t, Y_t)$ \label{line:pol}
	\State $O_t \gets \texttt{test}(V_t)$ \label{line:test}
	\State $ S_t \gets S \setminus O_t, \ A_t \gets A \setminus O_t, \ P_t \gets P \setminus O_t, \ Y_t \gets Y \setminus O_t$ \label{line:upd1}
	\State $\texttt{isolate}(O_t)$
	\label{line:upd2}
	\EndFor
\end{algorithmic}
\end{algorithm}

In Algorithm~\ref{alg:env}, we provide a high level description of the epidemic evolution in the case a testing policy is performed.
We also include the mechanisms regulating a generic tracing app, which will be used in the following sections to design our privacy-preserving test policy.
We are given with the initial populations $S_0$, $A_0$, $P_0$, $Y_0$, and $R_0$ and a time horizon $T$ for which we apply a testing policy $\texttt{testingPolicy}(\cdot)$, \emph{i.e.,} a strategy to select the individuals to test to limit the spread, and a swab test $\texttt{test}(\cdot)$ to check for the positivity to SARS-CoV-2.
At first, we initialize an empty set $D_u$ for each individual $u$, which will contain the information stored by the devices using the contact-tracing app during the period $\{1, \ldots, T\}$.
At each time $t$, we determine the new contacts $c_{u,v}(t) \in C(t)$ between individuals pairs, and we store the corresponding data on the devices $D_u$ and $D_v$ of the two individuals $u$ and $v$, respectively.
For the contact $c_{u,v}(t) \in C(t)$, it should be determined if a contagion between the involved nodes occurred (Line~\ref{line:cont}), and update the populations $S_t$, $A_t$, $P_t$, $Y_t$, and $R_t$ depending if the contagion generates a new asymptomatic or presymptomatic infection, or if some individuals recovered from the disease.
Finally, the policy $\texttt{testingPolicy}(\cdot)$ takes as input the contacts occurred in the past $C(1:t)$, the current population $U_t$, the newly symptomatics $Y_t$, and returns the individuals to be tested (Line~\ref{line:pol}).
Once the individuals $V_t$ are tested, populations $S_t$, $A_t$, $P_t$, $Y_t$ are updated, isolating positive individuals $O_t$ and those who manifested symptoms $Y_t$ (Lines~\ref{line:upd1}-\ref{line:upd2}).

\begin{algorithm}[t!]
\centering
\small
\caption{\texttt{contagionStep}($C(t), S_t, A_t, P_t, Y_t, R_t$)}
\begin{algorithmic}[1]
	\State Input: new contacts set $C(t)$, population sets $S_t$, $A_t$, $P_t$, $Y_t$, $R_t$
	\State Output: updated populations sets $S_t$, $A_t$, $P_t$, $Y_t$, $R_t$
	\For{$c_{u,v}(t) \in C(t)$}
		\If {$u \in S_t \wedge v \in \{A_t \cup P_t \cup Y_t \}$ and a contagion occurred (with probability $p(u, v, r_{u,v}(t), l_{u,v})(t)$)} \label{line:prob}
			\State $S_t \gets S_t \setminus \{ u \}$
			\If{ the contagion is asymptomatic (with probability $\alpha_s(u)$)}
				\State $A_t \gets A_t \cup \{ u \}$
			\Else
				\State $P_t \gets P_t \cup \{ u \}$
			\EndIf
		\EndIf
	\EndFor
	\For{$u \in P_t$}
		\If{$u$ expressed some symptoms}
			\State $P_t \gets P_t \setminus \{ u \}$
			\State $Y_t \gets Y_t \cup \{ u \}$
		\EndIf
	\EndFor
	\For {$u \in A_t \cup Y_t$}
		\If{$u$ recovered}
			\State $A_t \gets A_t \setminus \{ u \}$
			\State $Y_t \gets Y_t \setminus \{ u \}$
			\State $R_t \gets R_t \cup \{ u \}$
		\EndIf
	\EndFor
\end{algorithmic}
\label{alg:contagion}
\end{algorithm}

The evolution of the population sets describing the virus evolution is detailed in Algorithm~\ref{alg:contagion}.
At first, if a contact in $C(t)$ generated a new infected individual $u$, occurring with  probability $\delta(u, v, l_{u,v}(t), r_{u,v}(t))$ (Line~\ref{line:prob}), she/he is removed from the susceptible population $S_t$.
According to the response of the individual, \emph{i.e.,} she/he becomes a presymptomatic or an asymptomatic patient, she/he is moved from the susceptible set $S_t$ to either the set $P_t$ or $A_t$, respectively.
After that, if any presymptomatic $u \in P_t$ showed some symptoms, she/he is moved to the symptomatic set $Y_t$.
Finally, the contagion evolves by moving the status of all the asymptomatics and symptomatics who recovered to $R_t$.

\paragraph{Contact-tracing Model}
We provide a brief description on the contact tracing apps functioning in Algorithm~\ref{alg:devices}.
As mentioned before, even though in some countries these tools collect fine grained data including, beyond contacts, personal and GPS information, most of the European countries developed solutions that aim also at preserving privacy, by introducing some constraints on how data are collected.
These constraints brought the development of different solutions that exploit Bluetooth technologies and security protocols to preserve users privacy.
In what follows, we focus on the general schema followed by \emph{Immuni}, the official Italian contact tracing application, whose features are common with by most of the other European privacy-aware contact tracing applications.
The data update procedure generates for each contact $c_{u,v}(t)$ two random codes, a.k.a.~token, $h_{u, t}$ and $h_{v, t}$, one for each individual involved in a contact (Lines~\ref{line:token1}-\ref{line:token2}).
Subsequently, it updates the data stored in each individual's device $D_u$ and $D_v$, adding an entry with the time $t$ the contact occurred, the tokens of the two individuals $h_{u,t}$ and $h_{v,t}$, the distance $l_{u,v}(t)$, and duration $r_{u,v}(t)$ of the contact (Lines~\ref{line:d1}-\ref{line:d2}).
%

\begin{algorithm}[t!]
	\centering
	\small
	\caption{\texttt{updateDevData}($D_u, D_v, c_{u, v}(t)$)}
	\begin{algorithmic}
		\State Input: devices data $D_u$, $D_v$, new contact $c_{u, v}(t)$
		\State Output: updated device data $D_u$, $D_v$
		\State $h_{u,t} \gets $ \texttt{RandomToken}() \label{line:token1}
		\State $h_{v,t} \gets $ \texttt{RandomToken}() \label{line:token2}
		\State $D_u \gets D_u \cup \{(t, h_{u,t}, h_{v,t}, l_{u,v}(t), r_{u,v}(t)) \}$ \label{line:d1}
		\State $D_v \gets D_v \cup \{(t, h_{v,t}, h_{u,t}, l_{u,v}(t), r_{u,v}(t)) \}$ \label{line:d2}
	\end{algorithmic}
	\label{alg:devices}
\end{algorithm}

\section{Problem Formulation}

Given a population $U_0$ and a time horizon $T \in \mathbb{N}$, ideally, we aim at finding, at each day $t \in \{1, \ldots, T\}$, a subset $X_t \subset U_t$ of $K$ nodes to isolate to minimize the cumulative number of infected over the time horizon $T$.
%
%
However, solving such a problem would require full knowledge of the contacts probabilities $w(u, v)$, as well as the outcomes (contagion/no contagion) of each contact $c \in C(t)$.
Unfortunately, in our scenario, this information is not available due to the large number of asymptomatic and presymptomatic individuals, who do not allow to determine with certainty if a contagion occurred as a consequence of a contact.
Moreover, the classical analysis of the contagion chains cannot be performed in a centralized way: in our setting the privacy policies of the contact tracing apps do not allow the storage, not even temporary, in a centralized manner.
Indeed, we are able only to exploit the data coming from the different devices $D_u$ in a decentralized fashion.
%
%
In particular, we are allowed only to use the data stored in the device database $D_u$ and a model computing the probability of getting infected for each contact between individuals $\delta(u, v, l_{u,v}(t), r_{u,v}(t))$.
Therefore, we restate the problem of minimizing the virus spread as the task of finding, at every time $t$, the subset of nodes $X_t \in U_t$, with cardinality $K$, having largest probability $p^c_t(u)$ of being infected at $t$.
Furthermore, assuming that an individual can be isolated if she/he has evident symptoms or after resulting positive from a test, and that the symptomatic individuals are known, the problem can be formally cast as the one of selecting the nodes with largest probability of being infected $p_t^c(u)$:
\begin{align}\label{eq:obj}
	X_t = &\arg \max_{X \in U_t} \sum_{u \in X} p^c_t(u)\\
	&\text{ s.t. } |X_t| = K,
\end{align}
where $|\cdot|$ denotes the cardinality operator.
%

\section{The \alg{} Algorithm}

For the sake of simplicity, our algorithm is based on two assumptions: there is no environmental source of contagion, and all the individuals that present symptoms or resulted to be positive to a test communicate their status to the local health authorities.
While this is not the case in practice, the algorithm can be easily extended to capture these features. 
The key idea of our algorithm, namely Privacy Preserving Test Optimization (\alg{}), is to use a distributed/decentralized scheme to find the solution to the problem in Equation~\ref{eq:obj}.
Indeed, a distributed computation of the solution satisfies the privacy requirements imposed by the privacy-preserving policy of the governments.
Starting from the devices of symptomatic (positive) individuals $Y_t$, we aim at identifying asymptomatic $A_y$ and presymptomatic $P_t$ individuals by reconstructing infection trajectories, \emph{i.e.}, the paths of contacts the infection followed, exploiting the contact-tracing apps data.
The algorithm assigns to each user a score representing the infection risk level by recording how many times a Monte Carlo-simulated infection trajectory passed trough her/his device.
Finally, given a limited amount of $K$ available tests, the algorithm to swab test the set $X_t$ of the $K$ individuals with the highest score.

\begin{algorithm}[t!]
\centering
\small
\caption{\alg{} - Main}
\begin{algorithmic}
	\State Input: number of available tests $K$, number of iterations $N$, current time step $t$, time window $t_w$, set of newly positive discovered individuals $Q_t$, tokens of the contacts for the positive individuals $\{ H_u \}_{u \in Q_t}$
	\State Output: individuals to test $X_t$
	\For{$n \in \{1, \ldots, N \}$}
	\State Sample randomly an individual $u \in Q_t$ and $h \in H_u$
	\State \texttt{sendRequest}($n, h$)
	\EndFor
	\State $S = \emptyset$
	\While{StopCondition=False}
	\State $s =\texttt{ListenScores()}$
	\State $S = S \cup s$
	\EndWhile
	\State $X_t = \texttt{selectTopK}(S, K)$
	\State \texttt{sendNotifications}($X_t$)
\end{algorithmic}
\label{alg:pato-main}
\end{algorithm}

In Algorithm~\ref{alg:pato-main}, we provide the pseudo-code of our algorithm.
The \alg{} algorithm takes as input the number of available tests $K$, the number of daily Monte-Carlo iterations $N$, the current time $t$, and a time window $t_w$ for which we have to look for contagious people during the past days.\footnote{
In principle, one can use all the contacts stored in the devices $C(1:t)$, but for both computational reasons and given by medical evidence that $\tau_u \leq 14$~\cite{who2019media} and $\epsilon_u \leq 14$~\cite{lauer2020incubation}, a reasonable choice is to use $C(t-t_w:t) :=  \cup_{h=t-t_w}^t C(h)$ with $t_w = 14$.}
Notice that the use of $N$ is required to limit the computational effort of the algorithm we run.
Moreover, the \alg{} algorithm requires the set of individuals $Q_t := Y_t \setminus Y_{t - t_w}$ that recently, \emph{i.e.}, in the last $t_w$ days, were positive to the clinical tests and then have been communicated to the authorities together with their set of tokens used in the recent past, \emph{i.e.}, we need $H_u := \{ h_{v,j} \in D_u \text{ s.t. } j \in \{ t - t_w, \ldots, t \} \}$ for every individual $u \in Q_t$. 
Then, for $N$ iterations, the algorithm samples a positive infected node $u \in Q_t$ and simulate contagion trajectories starting from her/him, by selecting a token $h \in H_u$.
A simulation is started by broadcasting a request \texttt{sendRequest}($n, h, t$) targeting the device associated to the tokens $h$, and the current algorithm iteration $n$.
After a stop condition is met, \emph{e.g.,} a prespecified time has passed from the last iteration started, the algorithm listens to the device communicating in an anonymous way their scores $s = (x, g)$, composed by a device identifier $x$ and a contagion risk $g$.
Finally, using the values of the contagion risk, it communicates to the $K$ individuals with the largest score they are the candidates for the swab tests.

\begin{algorithm}[t!]
\centering
\small
\caption{\alg{} - Device Side }
\begin{algorithmic}
	\State \textbf{Input}: number of iterations $N$, starting time step $t$, time window $t_w$
	\State \textbf{Output}: score $g$
	\State Initialize score $g \gets 0$
	\State Initialize $f_j \gets False, \forall j \in \{1, \ldots, N\}$
	\While{Stop Condition = False}
		\State $(n, h) \gets$ \texttt{waitRequest}()
		\If{ $\exists d = (t', h_{u,t'}, h_{v,t'}, l_{u,v}(t'), r_{u,v}(t')), d \in D_u \mid h_{v,t'} = h \wedge f_n = False $}
			\State $f_n \gets True$
			\State $g \gets g + 1$
			\State \texttt{backwardTrajectory}($t', n$)
			\State \texttt{forwardTrajectory}($t', n$)
		\EndIf
	\EndWhile
	\State $x = \texttt{randomCode()}$
	\State $s = (x,g)$
	\State \texttt{sendScore}($s$)
\end{algorithmic}
\label{alg:pato-device}
\end{algorithm}

In Algorithm~\ref{alg:pato-device}, we provide the operations a device performs once it receives a request.
More specifically, each device waits for the requests sent through the network, but only if the token $h$ corresponds to one of those generated by the device, it will react.
If the token matches the device, then the device score $g$ is incremented, and a flag $f_n$ associated to the current iteration $n$ is raised.
After that, the device propagates the potential infection backward and forward in time to other individuals, as described in the procedures detailed in Algorithm~\ref{alg:pato-backward} and Algorithm~\ref{alg:pato-forward}, respectively. 
Finally, once a stop condition is satisfied, each device $u$ communicates its score $s$ in an anonymous way to a central unit (represented here by the core procedure in Algorithm~\ref{alg:pato-main} of the \alg{} algorithm).
Using Algorithm~\ref{alg:pato-backward}, the contagion trajectory is propagated by the device of individual $u$ in the past, \emph{i.e.,} to the devices which had a contact with $u$ during the period $\mathcal{B} = \{ t - t_w, \ldots, t' - 1\}$.
The device selects all the contacts occurred during the above mentioned period, formally defined as:
\begin{equation} \label{eq:back}
	\hspace{-2pt} C^B \hspace{-4pt} := \{ (j, h_{u,j}, h_{v,j}, l_{u,v}(j), r_{u,v}(j)) \hspace{-2pt} \in \hspace{-2pt} D_u, \text{s.t. } j \hspace{-2pt} \in \hspace{-2pt} \mathcal{B} \},
\end{equation}
and selects one element $c$ sampling it from $C^B$ according to a probability mass function:
\begin{align}
	\gamma(c) & = \left[ \prod_{c' \in C^B \; : t'' < t'} \hspace{-12pt} (1 - \hat{\delta}(u, v, l'_{u,v}(t''), r'_{u,v}(t''))) \right] \cdot \nonumber\\ &\hspace{20pt} \hat{\delta}(u, v, l_{u,v}(j), r_{u,v}(j)), \label{eq:gamma}
\end{align}
where $c' = (t'', h_{u,t''}, h_{v,t''}, l'_{u,v}, r'_{u,v})$ and
\begin{equation*}
	\hat{\delta}(u, v, l'_{u,v}, r'_{u,v})) = \hspace{-10pt} \sum_{B \in \{A_t, P_t, Y_t\}} \hspace{-10pt} \mathbb{P}(B) \cdot \delta(u, v, l'_{u,v}, r'_{u,v}) \big|_{v \in B},
\end{equation*}
is the probability that $v$ infected $u$, conditioned by the population of $v$ and weighted by the probability of belonging to that population, with $\mathbb{P}(B) := \frac{|B|}{|A_t \cup P_t \cup Y_t|}$.
This function represents an unnormalized probability that a node is infected by $v$, given that she/he has not been infected from the previous contacts, and is defined using the duration, the distance, and the order in which each contact occurred.
Finally, a single contact $\hat{c} = (\hat{t}, h_{u,\hat{t}}, h_{v,\hat{t}}, \hat{l}_{u,v}(\hat{t}), \hat{r}_{u,v}(\hat{t}))$ is selected and a request is sent for the token $h_{v,\hat{t}}$.
Conversely, using Algorithm~\ref{alg:pato-forward}, the contagion trajectory is propagated in the future, \emph{i.e.}, to the devices which had a contact during the period $\mathcal{F} = \{ t' + 1, \ldots, t\}$.
Once we selected all the contacts:
\begin{equation} \label{eq:forw}
	\hspace{-2pt} C^F \hspace{-4pt} := \hspace{-2pt} \{ (j, h_{u,j}, h_{v,j}, l_{u,v}(j), r_{u,v}(j)) \hspace{-2pt} \in \hspace{-2pt} D_u, \text{s.t. } j \hspace{-2pt} \in \hspace{-2pt} \mathcal{F} \},
\end{equation}
the algorithms simulates the contagion, \emph{i.e.,} the individual $u$ infects the individual $v$ with probability $p_c(l_{u,v}(j), r_{u,v}(j))$.
If for a specific contact $\hat{c} = (\hat{t}, h_{u,\hat{t}}, h_{v,\hat{t}}, \hat{l}_{u,v}(\hat{t}), \hat{r}_{u,v}(\hat{t}))$ the contagion occurred, the current device sends a request with the corresponding token $\hat{h}_{v, \hat{t}}$.

\begin{algorithm}[t!]
	\centering
	\small
	\caption{\texttt{backwardTrajectory}($t', n$) }
	\begin{algorithmic}
		\State \textbf{Input}: starting time $t'$, iteration index $n$
		\State Select contacts $C^B$ as in Equation~\eqref{eq:back}
		\For{$c \in C^B$}
			\State Compute $\gamma(c)$ as in Equation~\eqref{eq:gamma}
		\EndFor
		\State Sample a contact $\hat{c} = (\hat{t}, h_{u,\hat{t}}, h_{v,\hat{t}}, \hat{l}_{u,v}(\hat{t}), \hat{r}_{u,v}(\hat{t}))$ from $C^B$ according to the distribution $\frac{\gamma(c)}{\sum_{e \in C^B} \gamma(e)}$
		\State \texttt{sendRequest}($n, h_{v,\hat{t}}$)
	\end{algorithmic}
	\label{alg:pato-backward}
\end{algorithm}

\begin{algorithm}[t!]
	\centering
	\small
	\caption{\texttt{forwardTrajectory}($t', n$) }
	\begin{algorithmic}
		\State \textbf{Input}: starting time $t'$, iteration index $n$
		\State Select contacts $C^F$ as in Equation~\eqref{eq:forw}
		\For{$c \in C^F$}
			\State simulate a contagion with probability $p_c(l_{u,v}(j), r_{u,v}(j))$
			\If{the contagion occurred for $\hat{c} = (\hat{t}, h_{u,\hat{t}}, h_{v,\hat{t}}, \hat{l}_{u,v}(\hat{t}), \hat{r}_{u,v}(\hat{t}))$}
				\State \texttt{sendRequest}($n, h_{v,\hat{t}}$) 
			\EndIf
		\EndFor
	\end{algorithmic}
	\label{alg:pato-forward}
\end{algorithm}

\begin{figure*}[th!]
	\hspace{-0.4cm}
	\subfloat[]{\includegraphics[width=0.35\linewidth]{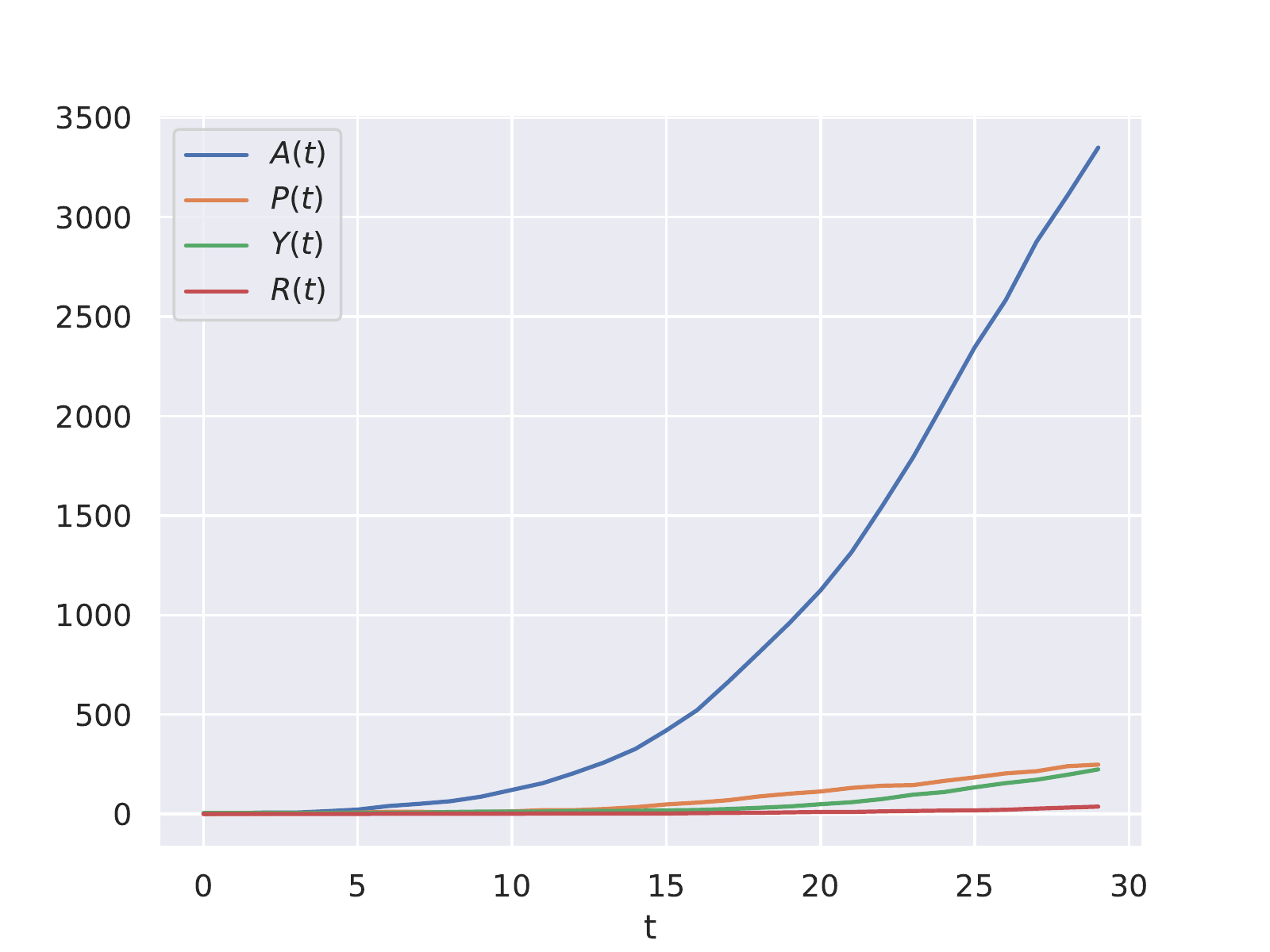}}
	\subfloat[]{\includegraphics[width=0.35\linewidth]{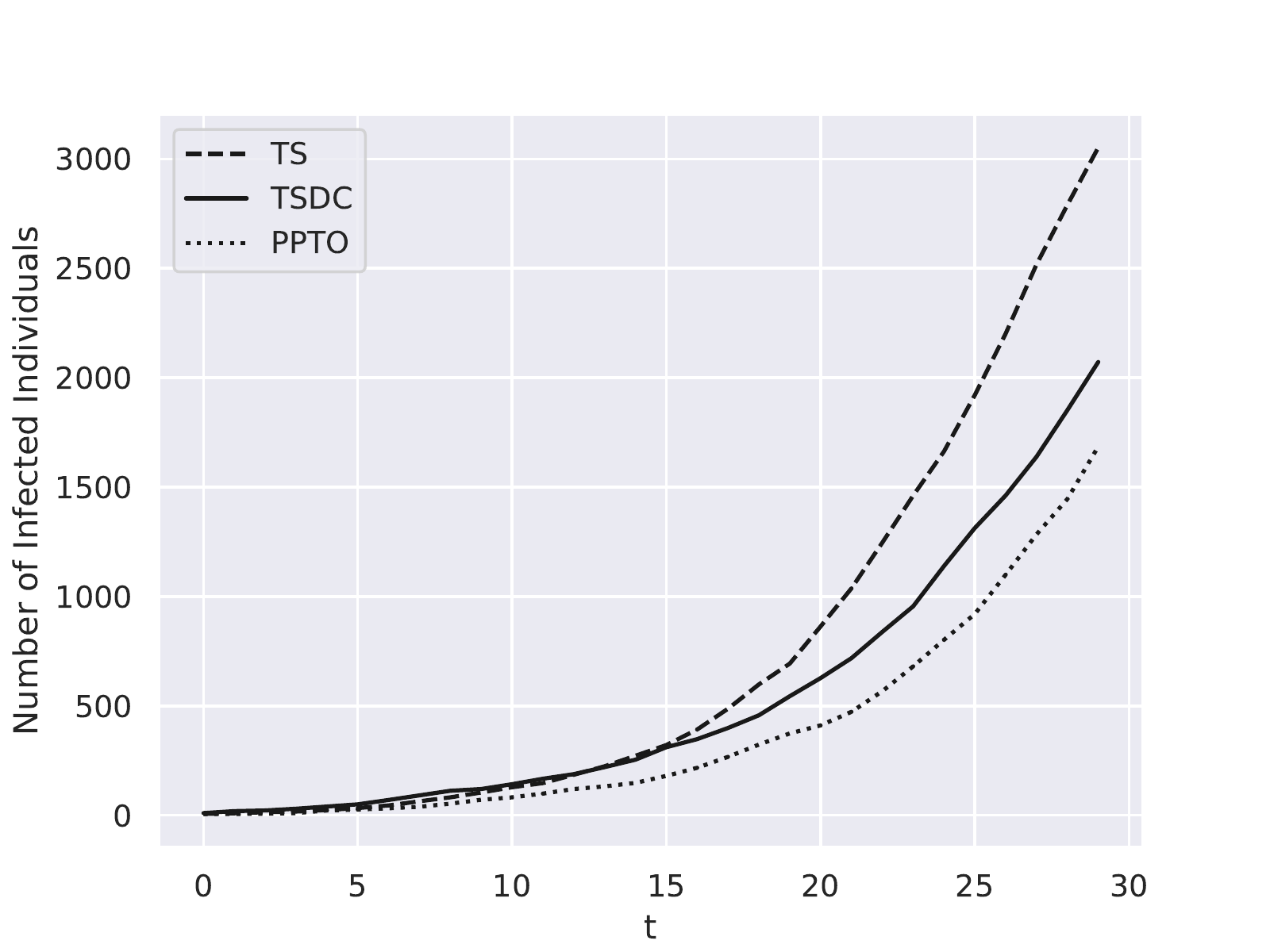}}
	\subfloat[]{\includegraphics[width=0.35\linewidth]{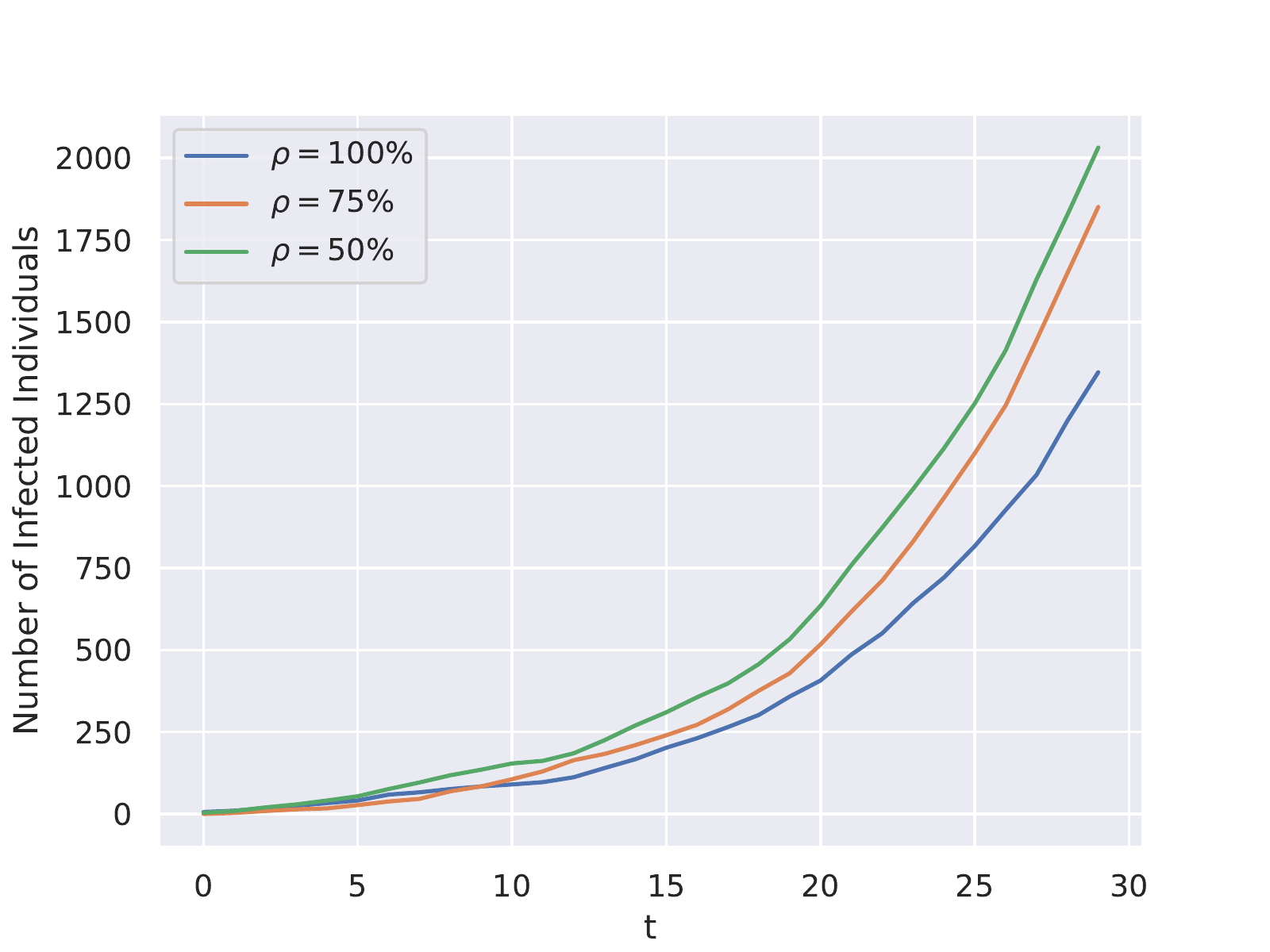}}
	\caption{(a) Values of infected individuals $|A_t|, |P_t|, |Y_t|$, and recovered ones $|R_t|$ without any containment measure; (b) number of infected individuals $|A_t|+|P_t|+|Y_t|$ appling \textsf{TS}, \textsf{TSDC}, and \alg{} testing policies; (c) number of infected individuals adopting \alg{} depending on the devices usage percentage $\rho$.}
	\label{fig:experiments}
\end{figure*}

\section{Experimental Evaluation}
In this section, we present an empirical evaluation of our algorithm in different settings generated by a simulator, and we compare its performance with the ones obtained with two baseline policies: \textsf{TS} tests all and only the newly symptomatics individuals, and \textsf{TSDC} tests all the newly symptomatics individuals and their direct contacts.
With both policies, if the available $K$ tests are not entirely used, the remaining tests are used randomly over the remaining susceptible individuals.
We compare the policies in terms of cumulative number of newly infected individuals during a time horizon $T = 30$ days.

\paragraph{Experiment \#1}

We have an initial set of $|U_0| = 10,000$ individuals.
The contact probability $w_{u,v}$ is set s.t., for each individual $u \in U_t$, the average number of daily contacts is $10$, the initial number of infected nodes to $|Y_0| = 5$, and the number of daily available tests to be $K = 100$.
In this experiment, for sake of simplicity, we sample $\tau_u$ from a uniform distribution $\mathcal{U}(5, \ 15)$, and $\epsilon_u$ from a uniform distribution $\mathcal{U}(1, \ 12)$.\footnote{
We report in Table~\ref{tab:settings1}, in the Supplementary Material, the value of the parameters used for the epidemic model.
}
%
%
We set the \alg{} algorithm to perform $N = 100$ iterations each day.
In Figure~\ref{fig:experiments}(a), we show the evolution populations $|A_t|$, $|P_t|$, $|Y_t|$, $|R_t|$ without any containment measure, \emph{i.e.}, in a setting with $K = 0$. 
We notice that after $30$ days, the asymptomatic, presymptomatic, and symptomatic populations reach the $33\%$, $2.48\%$, and $2.3\%$ of the whole population, respectively, while only the $0.35\%$ of the individuals have recovered from the disease.
In Figure~\ref{fig:experiments}(b), we show how the number of infected individuals grows depending on the testing policy we adopt. 
Even if no policy stops completely the virus spread in these settings, we notice significant differences the virus spread induced by the three policies.
More precisely, the adoption of the \alg{} policy reduces of the number of infected of $50\%$ compared with the \textsf{TS} policy, while it determines a reduction of the $20\%$ of the infected individuals compared with the \textsf{TSDC} policy.
This suggests that a more intelligent use of the available test might reduce the virus spread in a significant way.

\paragraph{Experiment \#2}
The goal of this experiment is to show that the collaboration of the users is crucial to contain the virus spread.
We evaluate the performance of our approach depending on the app usage percentage $\rho$ among the population $U_0$.
By varying this value, we determine the percentage of times in which the devices are actives and, therefore, the contacts are stored.
We use the parameters settings described in Table~\ref{tab:settings2} of the Supplementary Materials and we evaluate the \alg{} algorithm with different values of $\rho \in \{100\%, 75\%, 50\%\}$.
In Figure~\ref{fig:experiments}(c), we show the performance in terms of the number of the infected individuals depending on the value of $\rho$.
In this setting, a full collaboration of the individuals significantly affects the spread development.
More specifically, we note that by reducing the usage frequency to the $75\%$ we increase the spread to the $28\%$, while a value $\rho = 50\%$ would increase of the $35\%$ the number of infected at the end of the time horizon $T$.

\paragraph{Technological Considerations}
The \alg{} algorithm can be easily implemented in a decentralized fashion.
Indeed, every device propagates the information on the simulation at most once per run, and the time needed for this propagation is short, thus requiring a negligible computational effort for every device. 
Furthermore, Monte Carlo simulation provides a good approximation of the score with a reasonable number of runs.
This is well-known in the social influence maximization field~\cite{DBLP:conf/aaai/OhsakaAYK14} that is strictly related to our problem, and it is confirmed by our experimental activities (see Supplemental Material for the details).
\section{Discussion and Future Works}
In this paper, we present a novel algorithm for the identification of individuals that have high probability to spread the infection during the SARS-CoV-2 outbreak.
We show how to use the contact tracing apps and the data stored therein to provide with a suggestion on the individuals to test for the positivity to the virus.
This is crucial task for the containment of the contagion, due to the limited number of available tests the public health organizations can perform each day.
Our algorithm suggests the set of individuals that should be tested to limit the spread by an innovative AI-based algorithm which communicates in a decentralized fashion only by means of token generated by the different devices, thus preserving the private information of the app users.
As immediate future work, we will evaluate our algorithm in realistic scenarios comparing it with testing policies currently employed by governments.
However, simulating realistic scenarios is a hard task since the actual values of some parameters of our epidemic model are still unknown to the scientific community, such as the probability of being asymptomatic, the contagion probability given a contact.
For this reason, in addition to resort to new works in the scientific literature, we will design methods to estimate them by combining aggregated data about this epidemics.
Furthermore, another interesting improvement will be the development of a framework in which these parameters can be estimated online gathering information trough the tests.



\clearpage

\bibliographystyle{named}
\bibliography{refs}

\clearpage
\section{Supplementary Material (Paper ID: 5708)}

\begin{table}[H]
	\caption{Experiment \#3. Values of $R_T$ depending on the value of the number of iterations $N$.}
	\label{tab:exp3}
	\centering
	\begin{tabular}{lllll}
		\hline
		\multicolumn{1}{|c|}{}      & \multicolumn{1}{c|}{$N=50$} & \multicolumn{1}{c|}{$N=100$} & \multicolumn{1}{c|}{$N=500$} & \multicolumn{1}{c|}{$N=1000$} \\ \hline
		\multicolumn{1}{|c|}{$R_T$} & \multicolumn{1}{c|}{$0.0072$}      & \multicolumn{1}{c|}{$0.0172$}       & \multicolumn{1}{c|}{$0.0202$}       & \multicolumn{1}{c|}{$0.02103051$}        \\ \hline
		&                             &                              &                              &                               \\
		&                             &                              &                              &                              
	\end{tabular}
\end{table}
\subsection{Notation}

\begin{table}[H]
\small
\caption{Notation.}
\label{tab:notation}
\begin{center}
\begin{tabular}{r c p{5.4cm} }
	\toprule
	$T$ & $\triangleq$ & time horizon\\
	$U_t$ & $\triangleq$ & set of individuals involved in the contagion at day $t$\\
	$S_t$ & $\triangleq$ & susceptible individuals at day $t$\\
	$A_t$ & $\triangleq$ & asymptomatic individuals at day $t$\\
	$P_t$ & $\triangleq$ & presymptomatic individuals at day $t$\\
	$Y_t$ & $\triangleq$ & symptomatic individuals at day $t$\\
	$R_t$ & $\triangleq$ & recovered individuals at day $t$\\
	$w_{u,v}$ & $\triangleq$ & probability a contact between $u$ and $v$\\
	$C(t)$ & $\triangleq$ & set of contacts occurred at day $t$\\
	$c_{u,v}(t)$ & $\triangleq$ & contact occurred between $u$ and $v$ at day $t$\\
	$\delta(u, v, l, r)$ & $\triangleq$ & probability of a contagion of $u$ by $v$\\
	$l_{u,v}$ & $\triangleq$ & distance of the contact between $u$ and $v$\\
	$r_{u,v}$ & $\triangleq$ & duration of the contact between $u$ and $v$\\
	$\alpha(u)$ & $\triangleq$ & probability of being asymptomatic for $u$\\
	$\tau_u$ & $\triangleq$ & recovering time for an asymptomatic individual\\
	$\epsilon_u$ & $\triangleq$ & number of days before symptoms occurs after a contagion\\
	$C(1:t)$ & $\triangleq$ & contacts occurred in the period $\{1, \ldots, t \}$\\
	$V_t$ & $\triangleq$ & set of individuals tested at day $t$\\
	$K$ & $\triangleq$ & number of available tests at day $t$\\
	$O_t$ & $\triangleq$ & set of individuals resulting positive to the test at day $t$\\
	$D_u$ & $\triangleq$ & data present on the device of individual $u$\\
	$h_{u,t}$ & $\triangleq$ & token sent by user $u$ during a contact event\\
	$p^c_t(u)$ & $\triangleq$ & probability of being infected for $u$\\
	\bottomrule
\end{tabular}
\end{center}
\end{table}
\begin{table}[H]
	\caption{Parameters of the synthetic settings adopted in Experiment 1.}
\label{tab:settings1}
\scriptsize
\centering
\begin{tabu}{|l|[2pt]l|}
	\hline
	Parameter & Value\\ \tabucline[2pt]{-}
	$|U|$       & 10000   \\  \hline
	$K$    & 100      \\  \hline
	$w_{u,v}$  & 0.001       \\  \hline
	$\delta(u \in S_t, v \in A_t,l = 0, r \in \{0,1\})$  & 0.02      \\  \hline
	$\delta(u \in S_t, v \in P_t,l = 0 ,r\in \{0,1\})$  & 0.05       \\  \hline
	$\delta(u \in S_t, v \in Y_t,l = 0, r \in \{0,1\} )$  & 0.07       \\  \hline
	$\delta(u \in S_t, v \in A_t,l= 1, r \in \{0,1\} )$  & 0.03      \\  \hline
	$\delta(u \in S_t, v \in P_t,l = 1,r \in \{0,1\} )$  & 0.06       \\  \hline
	$\delta(u \in S_t, v \in Y_t,l = 1, r \in \{0,1\} )$  & 0.08       \\  \hline
	$\alpha(u)$    & 0.1      \\  \hline
	$\beta(u \in A_t)$    & 0.4      \\  \hline
	$\beta(u \in P_t)$    & 0.2      \\  \hline	
	$\beta(u \in Y_t)$    & 0.05      \\  \hline
	$\tau_{min}$    & 5      \\  \hline
	$\tau_{max}$    & 15      \\  \hline
	$\epsilon_{min}$    & 1      \\  \hline
	$\epsilon_{max}$    & 12      \\  \hline
\end{tabu}
\end{table}

\begin{table}[H]
	\caption{Parameters of the synthetic settings adopted in Experiment 2.}
	\label{tab:settings2}
	\scriptsize
	\centering
	\begin{tabu}{|l|[2pt]l|}
		\hline
		Parameter & Value\\ \tabucline[2pt]{-}
		$|U|$       & 10000   \\  \hline
		$K$    & 100      \\  \hline
		$w_{u,v}$  & 0.07       \\  \hline
		$\delta(u \in S_t, v \in A_t,l=  \in \{0,1\}, r \in \{0,1\} )$  & 0.03      \\  \hline
		$\delta(u \in S_t, v \in P_t,l =  \in \{0,1\},r \in \{0,1\} )$  & 0.06       \\  \hline
		$\delta(u \in S_t, v \in Y_t,l =  \in \{0,1\}, r \in \{0,1\} )$  & 0.08       \\  \hline
		$\alpha(u)$    & 0.1      \\  \hline
		$\beta(u \in A_t)$    & 0.04      \\  \hline
		$\beta(u \in P_t)$    & 0.02      \\  \hline	
		$\beta(u \in Y_t)$    & 0.01      \\  \hline
		$\tau_{min}$    & 5      \\  \hline
		$\tau_{max}$    & 15      \\  \hline
		$\epsilon_{min}$    & 1      \\  \hline
		$\epsilon_{max}$    & 12      \\  \hline
	\end{tabu}
\end{table}

\subsection{Computing Infrastructure}
The code has been run on a Intel® Core™ $i5-6200U$ CPU @ 2.30GHz ($x4$) with $8$ GiB
of system memory. The operating system was Ubuntu $16.04$ LTS, and the experiments have been run on Python $3.5.2$ The libraries used in the experiments, with the corresponding version were:
\begin{itemize}
	\item numpy==1.17
	\item networkx==2.4
	\item seaborn==0.9.0
	\item matplotlib==3.1.3
\end{itemize}

\end{document}